# Probing magnetism in 2D van der Waals crystalline insulators via electron tunneling

D. R. Klein[1†], D. MacNeill[1†], J. L. Lado[2,3], D. Soriano[2], E. Navarro-Moratalla[4], K. Watanabe[5], T. Taniguchi[5], S. Manni[6-8], P. Canfield[6,7], J. Fernández-Rossier[2], P. Jarillo-Herrero[1*]

**Affiliations:**

[1]Department of Physics, Massachusetts Institute of Technology, Cambridge, MA 02139, USA.

[2]QuantaLab, International Iberian Nanotechnology Laboratory, 4715-310 Braga, Portugal.

[3]Institute for Theoretical Physics, ETH Zurich, 8093 Zurich, Switzerland.

[4]Institute of Molecular Science, University of València, València, Spain.

[5]National Institute for Materials Science, Tsukuba, Japan.

[6]Ames Laboratory, U.S. Department of Energy, Iowa State University, Ames, IA 50011, USA.

[7]Department of Physics and Astronomy, Iowa State University, Ames, IA 50011, USA.

[8]Department of Condensed Matter Physics and Materials Science, Tata Institute of Fundamental Research, Mumbai 400005, India.

[†]These authors contributed equally to this work.

[*]Correspondence to: pjarillo@mit.edu

**Abstract**: Magnetic insulators are a key resource for next-generation spintronic and topological devices. The family of layered metal halides promises ultrathin insulating multiferroics, spin liquids, and ferromagnets, but new characterization methods are required to unlock their potential. Here, we report tunneling through the layered magnetic insulator $CrI_3$ as a function of temperature and applied magnetic field. We electrically detect the magnetic ground state and inter-layer coupling and observe a field-induced metamagnetic transition. The metamagnetic transition results in magnetoresistances of 95%, 300%, and 550% for bilayer, trilayer, and tetralayer $CrI_3$ barriers, respectively. We further measure inelastic tunneling spectra for our junctions, unveiling a rich spectrum of collective magnetic excitations (magnons) in CrI$_3$. Our results establish vertical tunneling as a versatile probe of magnetism in atomically thin insulators.

**Main Text:**

Van der Waals magnetic insulators constitute a new materials system enabling designer topological states[1] and spintronic technologies[2]. The recent isolation[3,4] of few-layer magnets with either ferromagnetic (CrI$_3$, Cr$_2$Ge$_2$Te$_6$) or antiferromagnetic order[5,6], is just the tip of the iceberg. The vast family of layered metal halides[7] contains spin orders from multiferroics[8] to proximate spin liquids[9], of key interest to both fundamental and applied physics. However, this diversity requires new tools for exploring magnetism in atomically thin crystals. Existing studies have focused on magneto-optical effects[3,4,10,11], but a more general, device-oriented, approach is needed.

Here we introduce tunneling *through* layered insulators as a versatile probe of magnetism on the nanoscale. We report the conductance of graphite/CrI$_3$/graphite junctions (Fig. 1A) as a function of magnetic field and temperature, electrically detecting an antiferromagnetic ground state and a field-induced metamagnetic transition. The metamagnetic transition is revealed by large magnetoresistances (up to 550%) arising from the antiparallel to parallel reorientation of chromium spins in adjacent crystal layers. A similar effect was previously proposed[12] for synthetic multilayer magnets, but experimental realizations[13] were limited to magnetoresistances below 70%. The performance of our devices is an order of magnitude higher, corresponding to estimated spin polarization above 95%. Furthermore, the two-dimensional magnetism of CrI$_3$ enables ultra-thin tunnel barriers (< 3 nm) and a concomitant 10,000-fold increase in conductance (per unit area) compared to previous results[13]. Our devices therefore represent a new state-of-the-art in the electrical readout of insulating magnets, and the non-invasive van der Waals transfer of the magnetic layer ensures substrate independent device integration. This, together with high magnetoresistance, spin polarization, and conductance, will enable non-invasive spin injectors and detectors for next-generation spintronics experiments incorporating topological insulators[14], superconductors[15], antiferromagnets[16], and low symmetry crystals[17-20].

Tunneling through magnetic insulators was first studied in the pioneering experiments of Esaki[21] and later by Moodera *et al.*[22,23] When electrons tunnel through a ferromagnetic insulator, spin up and down electrons see different barrier heights (Fig. 1B). As a result, the tunneling rate can vary by orders of magnitude for electrons of opposite spins[12,22], called the spin filter effect. The smaller gap for spin up electrons tends to decrease the junction resistance as the barrier is cooled below its Curie temperature $T_C$. The situation is more complicated for spatially textured magnetism. For example, the resistance of Ag/EuSe/Al tunnel junctions *increases* significantly when the EuSe becomes antiferromagnetic[23]. However, in every case the exponential dependence of the tunneling current on the barrier electronic structure provides a clear resistive signature of magnetism. We will use these effects to electrically detect the magnetic ground state and field-induced metamagnetic transition of few layer $CrI_3$.

The resistance of a graphite/tetralayer $CrI_3$/graphite junction as a function of temperature is shown in Figure 1C. We measure the resistance in a four-point geometry using a 30 mV AC excitation[24]. The temperature dependence was measured by cooling the sample down with (purple line) and without (black line) the application of an external magnetic field. The magnetic field is applied perpendicular to the layers, along the magnetic easy axis of $CrI_3$. Above 90 K, the resistance is independent of the applied field and shows Arrhenius behavior with a thermal activation gap of roughly 159 meV (see Fig. S1). The resistance becomes field-dependent as the temperature approaches the bulk $T_C$ of 61 K. When the sample is cooled in a 2.5 T magnetic field, the resistance plateaus below 80 K, signaling the onset of tunneling conductance[21,23]. On the other hand, when the sample is cooled without an external field, the resistance exhibits a kink near $T_C$ and continues to increase below 60 K. The strong dependence of the tunneling resistance on magnetic field, as well as the temperature dependence, show that the tunnel conductance is sensitive to the magnetization of the barrier.

To further investigate the magnetic phase diagram, we study the zero bias conductance (500 µV AC excitation) of devices with two to four layer $CrI_3$ barriers as a function of applied magnetic field at low temperatures (300 mK to 4.2 K). We start with an analysis of a graphite/bilayer $CrI_3$/graphite junction (Fig. 2A). For this device, the junction conductance increases almost twofold in a sharp step as the external field is increased above 0.85 T. The corresponding magnetoresistance is 95%, defined as:

$$\text{MR} = 100 \times \frac{(G_{HI} - G_{LO})}{G_{LO}}, \tag{1}$$

where $G_{HI}$ is the high field conductance maximum and $G_{LO}$ is the low field conductance minimum. No further steps are observed up to the largest fields studied (8 T, see Fig. S2). As the field is reduced from 2.4 T, the conductance decreases to its original zero field value in a sharp step at 0.35 T. The well-defined steps and hysteretic field dependence demonstrate that the conductance changes originate from switching events of the magnetization. The tunneling current is most sensitive to the interlayer magnetization alignment, so the large steps we observe likely arise from *vertical* domains, *i.e.* regions where the magnetization points in different directions in different layers of $CrI_3$.

Recently, magneto-optical Kerr effect (MOKE) data has revealed an antiferromagnetic state in bilayer $CrI_3$ for fields below about 0.6 T[4]. In this state, the Cr moments order ferromagnetically within each layer, but point in opposite directions in adjacent layers (Fig. 2B).

The layers are fully aligned when the external magnetic field is increased above a critical value (Fig. 2C), *i.e.* it undergoes a metamagnetic transition to a ferromagnetic state. When the field is reduced, the magnetization spontaneously reverts to the antiparallel configuration. The switching behavior we observe in magnetoconductance reflects these previous MOKE data, confirming that the conductance change arises from the metamagnetic transition. We note, however, that little to no hysteresis was previously observed in the MOKE results for bilayer $CrI_3$[4], while we observe a clear hysteresis in the tunneling measurements. The reasons for this may be related to the lower temperature for these tunneling experiments as well as the fact that the $CrI_3$ remains closer to equilibrium (no photoexcitation).

We have also studied tunnel junctions with three and four layer $CrI_3$ as the barrier. The zero bias junction resistance of a graphite/4L $CrI_3$/graphite junction is shown as a function of external magnetic field in Fig. 2D. The overall phenomenology is similar to junctions with a bilayer barrier, with well-defined steps and a total magnetoresistance of 550%. The behavior of our trilayer junctions is again similar with magnetoresistances up to 300% (see Fig. S3). Based on these results, we hypothesize that few-layer $CrI_3$ is antiferromagnetic without an external magnetic field (Fig. 2E). Such behavior is consistent with magneto-optical data for bilayer $CrI_3$[4], but those MOKE data suggested a ferromagnetic configuration for thicker crystals (e.g. 3L $CrI_3$). Nevertheless, our data strongly supports an antiparallel alignment between layers extending over most of the junction area. Once more, the different temperatures and absence of photoexcitation may be responsible for the different behavior observed.

To understand the large magnetoresistance and its thickness dependence, we analyze a spin filter model[12] for transmission through a $CrI_3$ barrier. The model treats each crystal layer of the $CrI_3$ as an independent tunnel barrier, with a transmission coefficient of $T_P$ and $T_{AP}$ for spins parallel and antiparallel to the local spin direction, respectively. Ignoring multiple reflections and quantum interference effects, the transmission through entire crystal is then a product of the transmission coefficients for each layer. For example, for a CrI$_3$ bilayer in the high field magnetization configuration (Fig. 2C), spin up electrons have a transmission probability $T_P^2$ whereas spin down electrons have transition probability $T_{AP}^2$. The high field conductance is $G_{HI} \propto T_P^2 + T_{AP}^2$. Similarly, for the low field configuration with antiparallel magnetizations (Fig. 2B), the conductance is $G_{LO} \propto 2T_P T_{AP}$. The ratio of high field to low field conductances is then $G_{HI}/G_{LO}$ = $(T_P^2 + T_{AP}^2)/2T_P T_{AP} \approx T_P/2T_{AP}$. We have carried out similar calculations for $N$ = 3 and 4 layer CrI$_3$ barriers, summarized in the supplementary text. In Figure 3A, we plot the measured magnetoresistance (black circles) as a function of $N$, along with a one parameter fit to the spin filter model (purple stars). The model reproduces the overall experimental trend with a best fit value of $T_P/T_{AP}$ = 3.5.

We can also estimate the spin polarization of the current within the spin filter model. When the CrI$_3$ is fully polarized, the transmission probability of up and down spins though an $N$ layer CrI$_3$ barrier are $T_P^N$ and $T_{AP}^N$, respectively. Therefore, the ratio of spin up to spin down conductance is approximately $G_\uparrow/G_\downarrow = (T_P/T_{AP})^N$. From $T_P/T_{AP} \approx 3.5$, we estimate a spin polarization of $(G_\uparrow - G_\downarrow)/(G_\uparrow + G_\downarrow) \approx$ 85%, 95%, and 99% for N = 2, 3, and 4 respectively. These values are comparable to the largest values obtained with EuSe and EuS magnetic insulator barriers[13,23], so that CrI$_3$ tunnel barriers can enable future spin sensitive transport devices.

In the spin filter approximation, the calculation of the magnetoresistance is reduced to a calculation of $T_P/T_{AP}$, related to the different barrier heights for spin up and down electrons. To investigate the barrier heights, we carried out density functional theory (DFT) calculations for three layers of CrI$_3$ and three layers of graphite (see the supplementary text). Calculations portray

CrI$_3$ as a ferromagnetic insulator with magnetic moments localized on the chromium atoms and spin split energy bands (Fig. 3C). Importantly, when the magnetization of the three layers are aligned, we find that spin up bands of CrI$_3$ lie very close to the graphite Fermi energy, whereas the nearest spin down bands are much higher in energy (>1 eV). Therefore, the transparency of the barrier has to be much smaller for spin down electrons and provides a microscopic foundation for the large $T_P/T_{AP}$. Note that even though the DFT calculations show a CrI$_3$ majority band very close to or crossing the graphite Fermi energy, the exponential thickness dependence of the junction resistance (Fig. 3B) proves that our junctions are in the tunneling dominated regime with a finite barrier height. Further transport calculations should elucidate the precise tunneling pathways in CrI$_3$/graphite junctions leading to finite energy barriers with chromium 3$d$ orbital bands very close to the Fermi level.

In addition to the zero-bias conductance, we measured the differential conductance d$I$/d$V$ as a function of the applied DC offset $V_{DC}$. The d$I$/d$V$ versus $V_{DC}$ traces reveal a rich spectrum, whose most prominent features are a series of step-like increases, symmetric in bias, below 25 meV (Fig. 4, also Fig. S4 and Fig. S5). These steps are characteristic of inelastic electron tunneling where electrons lose energy to collective excitations of the barrier or electrodes. When the tunneling energy ($eV_{DC}$) exceeds the collective excitation energy, the introduction of these additional tunneling pathways results in steps in the d$I$/d$V$ versus $V_{DC}$ trace. The energies of phonons[25-27] and magnons[28-31] can therefore be measured as peaks (dips) in d$^2I$/d$V^2$ versus $V_{DC}$ for positive (negative) $V_{DC}$. The bottom panel of Fig. 4A shows |d$^2I$/d$V^2$| obtained by numerical differentiation of the d$I$/d$V$ data. The inelastic tunneling spectrum (IETS) reveals three peaks at 3 meV, 7 meV, and 17 meV. These peaks were visible in every CrI$_3$ tunneling device we measured. Past IETS data on graphite/boron nitride/graphite heterostructures in a similar geometry to our junctions[26] do not contain any inelastic contributions from graphite phonons below 17 meV. Earlier scanning tunneling studies of graphite surfaces reach similar conclusions[27]. Thus, the inner two peaks must arise from $CrI_3$ phonons or magnons. The inelastic features start forming just below the onset of magnetism (see Fig. S6), suggesting a magnon excitation origin.

Another signature of magnon-assisted tunneling is the stiffening of the magnon modes as an external magnetic field is applied[29,30]. A single magnon corresponds to a delocalized spin-flip within the $CrI_3$ barrier, which carries a magnetic moment of approximately $2\mu_B$ ($|S_z| = 1$) antiparallel to the external magnetic field. Therefore, magnon IETS peaks should blueshift at 0.12 meV/T by the Zeeman effect. Figure 4B shows $|d^2I/dV^2|$ as a function of both applied magnetic field and bias voltage. Even by eye, a strong linear increase of all three IETS peaks is visible. In Figure 4C we plot the peak energies (determined by Gaussian fits) of the innermost peaks versus magnetic field. We have also plotted the expected energy shift $2\mu_B B$ due to the Zeeman effect (dashed grey line). This line roughly fits the magnetic field dependence of the 3 meV peak, but the 7 meV peak clearly has much higher dispersion corresponding to 8 Bohr magnetons. The latter effect might be due to magnon renormalization effects, as discussed below.

To model the magnon spectrum, we write an effective spin Hamiltonian for $CrI_3$[32] that includes nearest and next-nearest neighbor exchange, together with an easy axis anisotropy term (see supplementary text for details). Using this model, we find that the calculated magnon density of states can qualitatively reproduce the experimental inelastic spectrum (Fig. 4D). At zero temperature, the magnon energies are still expected to blueshift at 0.12 meV/Tesla in an applied magnetic field (Fig. 4E). However, at finite temperature and $B=0$, thermally excited magnons deplete the magnetization, resulting in an effective reduction of the spin stiffness, and a red shift of the magnon spectrum. Application of a magnetic field increases the spin wave gap, decreasing the population of thermal spin waves and increasing the spin stiffness. This renormalizes the effective magnon hopping parameters, leading to a shift of the spin wave spectrum that adds to the Zeeman term and results in a nonlinear field dependence (Fig. 4F).

To summarize, we observe a large magnetoresistance in graphite/$CrI_3$/graphite spin filter devices corresponding to estimated spin polarization over 95% for trilayer and tetralayer $CrI_3$ barriers. The magnetoresistance ratios we observe, as high as 550%, are comparable to optimized industry magnetic tunnel junctions, and there is extensive room for improvement beyond our basic proof of concept devices. The large magnetoresistance arises from antiferromagnetic coupling between adjacent $CrI_3$ layers, which frustrates transport in the zero field magnetization configuration. This is an example of the "double spin filter" effect where a magnetic tunnel barrier with decoupled magnetic layers is used as a magnetic memory bit[12]. Our devices surpass previous limitations of double spin filters[13] owing to the unique decoupling of magnetic layers across the atomic scale van der Waals gap. This decoupling provides electrical readout of the $CrI_3$ magnetization state without additional ferromagnetic sensor layers, enabling facile detection of spin-orbit torques on layered magnetic insulators. Finally, our devices show strong inelastic tunneling features consistent with magnon excitation. Further exploration is required to understand the electron-magnon coupling in these devices and to potentially study bosonic topological matter in honeycomb ferromagnets[33,34].

**Acknowledgments:**

We thank V. Fatemi and Y. Cao for helpful discussions and assistance with measurements. This work was supported by the Center for Integrated Quantum Materials under NSF grant DMR-1231319 as well as the Gordon and Betty Moore Foundation's EPiQS Initiative through Grant GBMF4541 to PJH. Device fabrication has been partly supported by the Center for Excitonics, an Energy Frontier Research Center funded by the US Department of Energy (DOE), Office of Science, Office of Basic Energy Sciences under Award Number DESC0001088. DRK acknowledges partial support by the NSF Graduate Research Fellowship Program (GRFP) under Grant No. 1122374. JLL acknowledges financial support from the ETH Zurich Postdoctoral Fellowship program. DS acknowledges the Marie Curie Cofund program at INL. JFR thanks support from PTDC/FIS-NAN/3668/2014. Growth of hexagonal boron nitride crystals at NIMS was supported by the Elemental Strategy Initiative conducted by the MEXT, Japan and JSPS KAKENHI Grant Numbers JP15K21722 and JP25106006. The data presented in this paper are available from the corresponding authors upon reasonable request.

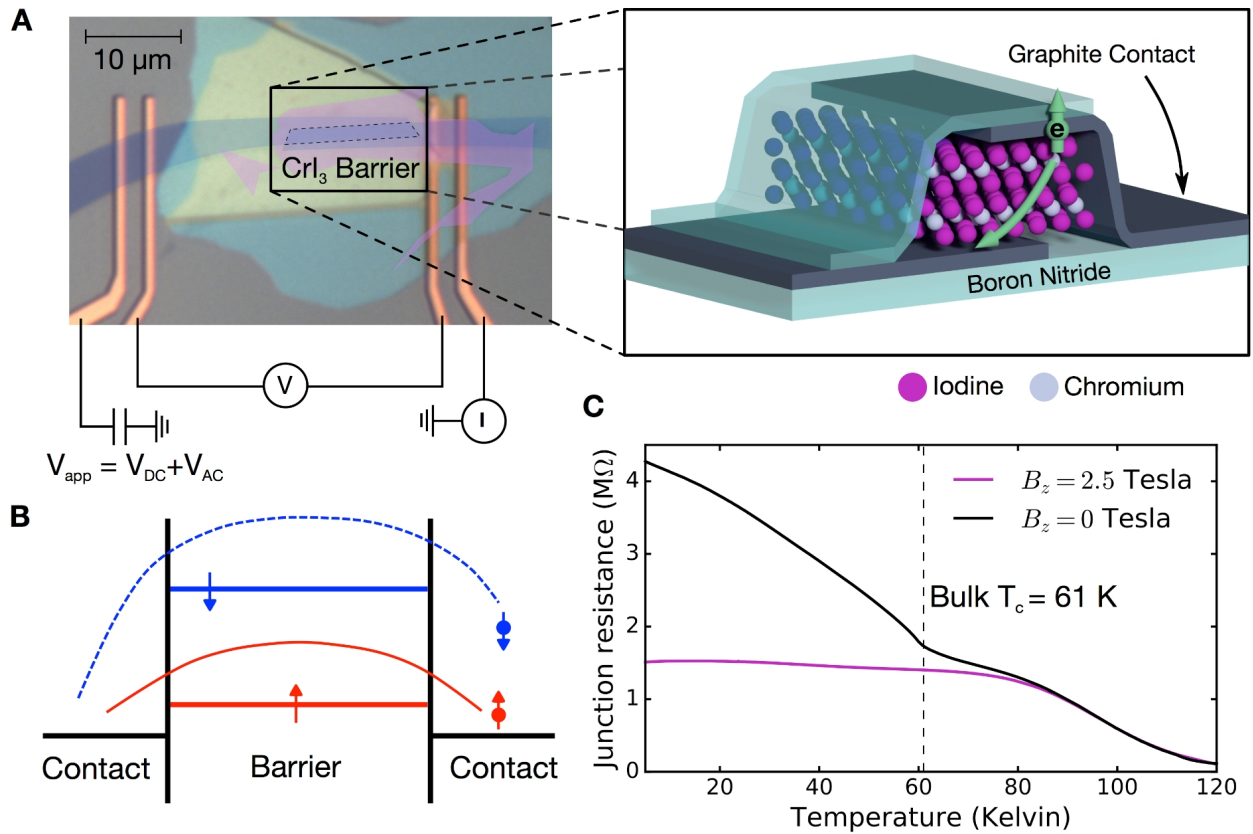

**Fig. 1. (A)** Optical micrograph of a tetralayer CrI$_3$ tunnel junction device (false color). The dashed line encloses the tunnel junction area. The graphite contacts are themselves contacted by Au/Cr wires in a four-point geometry. Inset: Schematic of the van der Waals heterostructures studied in this work. Electrons tunnel between two graphite sheets separated by a magnetic CrI$_3$ tunnel barrier. The entire stack is encapsulated in hexagonal boron nitride. **(B)** Schematic energy diagram of a metal/ferromagnetic insulator/metal junction. The red and blue lines in the barrier region represent the spin up and spin down energy barriers, respectively. The lower barrier for spin up electrons leads to spin polarized tunneling and reduced resistance for a ferromagnetic barrier. **(C)** Zero bias junction resistance vs. temperature for a graphite/tetralayer CrI$_3$/graphite junction cooled with (purple) and without (black) an applied magnetic field. The curves begin to deviate around

the bulk Curie temperature (61 K) giving evidence for magnetic order in the CrI$_3$ barrier and for spin polarized tunneling. The magnetic field was applied perpendicular to the CrI$_3$ layers.

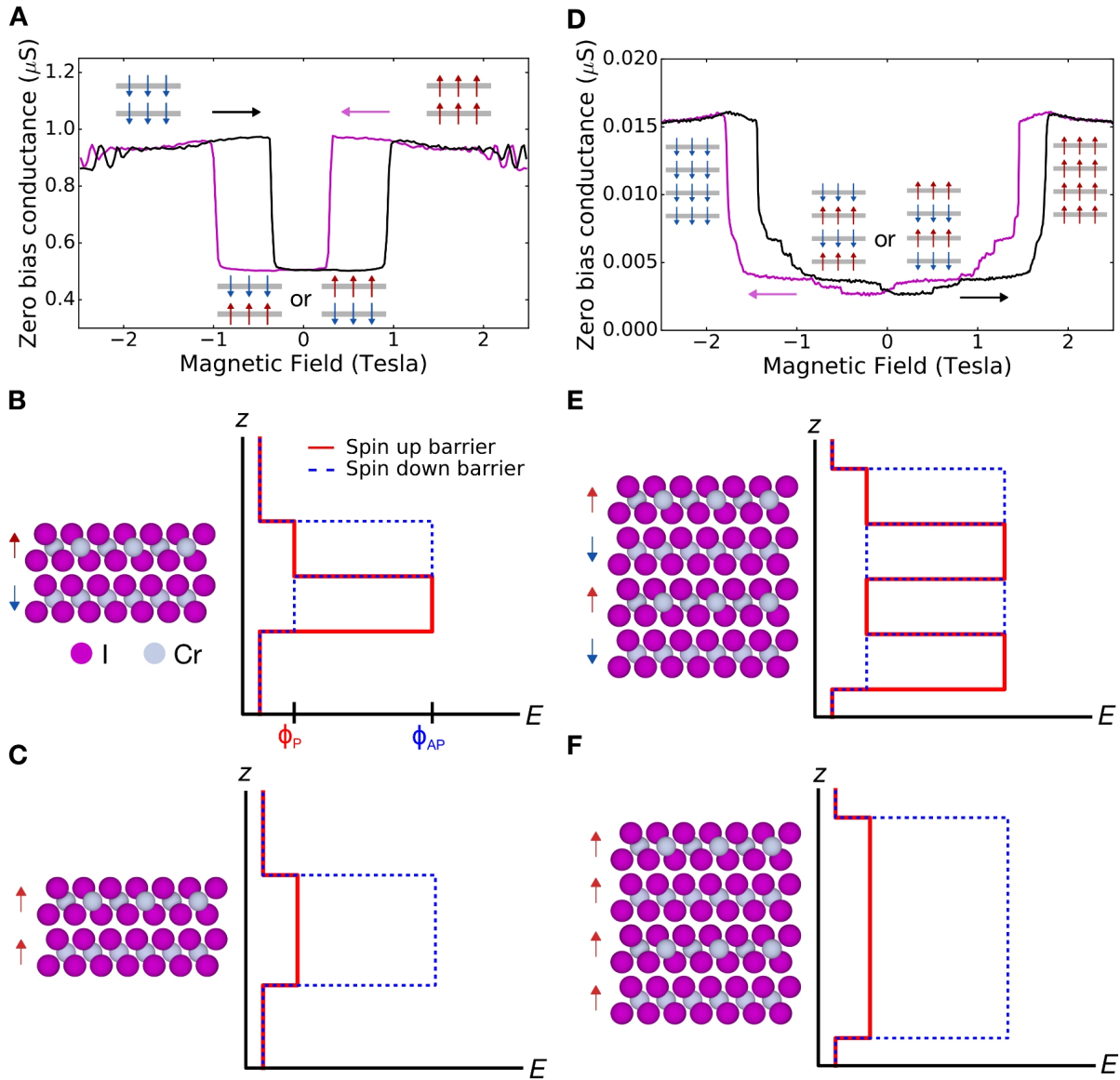

**Fig. 2. (A)** Conductance through a bilayer CrI$_3$ tunnel barrier as a function of an out-of-plane applied magnetic field with 500 µV AC excitation. The data was taken both for decreasing (purple line, left arrow) and increasing (black line, right arrow) magnetic field. The magnetoconductance traces match previous magnetometry data for bilayer CrI$_3$ showing that the two layers are antiparallel at zero field but can be aligned with an external field below 1 T. **(B, C)** Schematic of barriers experienced by spin up and spin down electrons tunneling through bilayer CrI$_3$ in the low and high field states. In the low field state, the two layers are antiparallel and both spins see a high barrier. In the high field state, the layers are aligned and up spins see a low energy barrier leading to increased conductance. **(D-F)** Analogous data and schematics for a tetralayer CrI$_3$ barrier device. In both cases, the sample temperature was 300 mK.

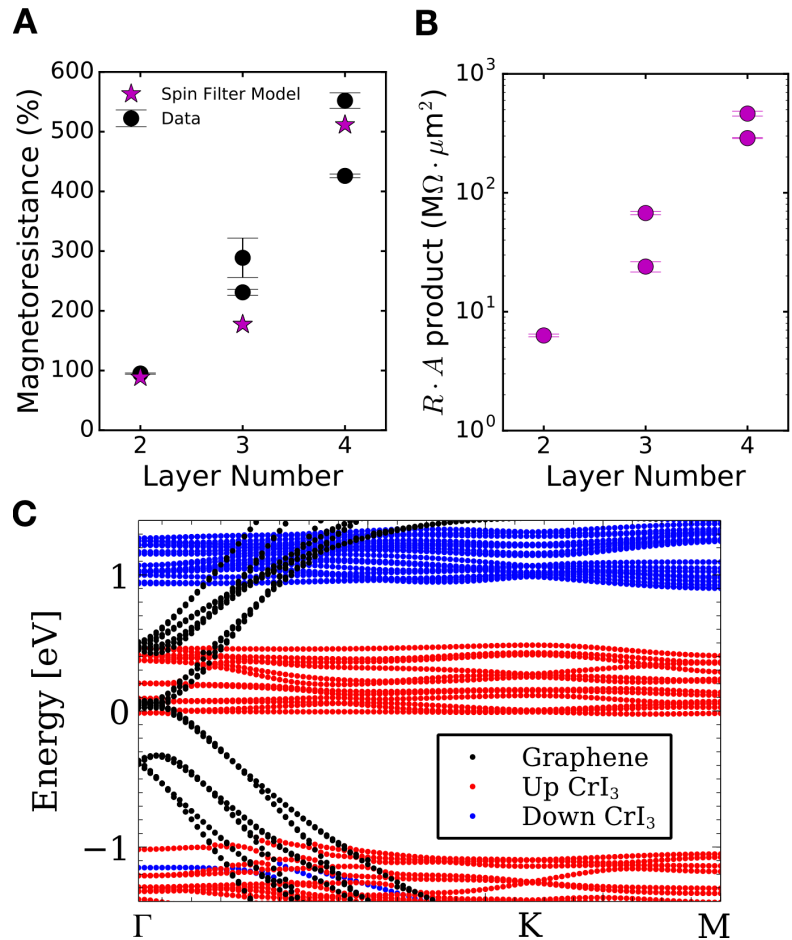

**Fig. 3. (A)** Magnetoresistance ratio (black circles) versus CrI$_3$ layer number for multiple devices. We also plot a fit to the spin filter model (purple stars). The only fitting parameter, $T_P/T_{AP} = 3.5$, gives the ratio of spin up to spin down transmission through a CrI$_3$ monolayer. **(B)** Resistance-area

product versus CrI$_3$ layer number for multiple devices. The resistances are measured in the fully aligned magnetic configuration and were taken at zero bias. **(C)** Electronic structure of a trilayer graphite/trilayer CrI$_3$ heterostructure calculated with density functional theory. The CrI$_3$ is in the fully ferromagnetic configuration and its bands are projected on the spin up and down channels. While the minority spins do not show states close to the Fermi energy, there is a large number of states in the majority channel. The difference establishes a microscopic basis for the large $T_P/T_{AP}$ we observe.

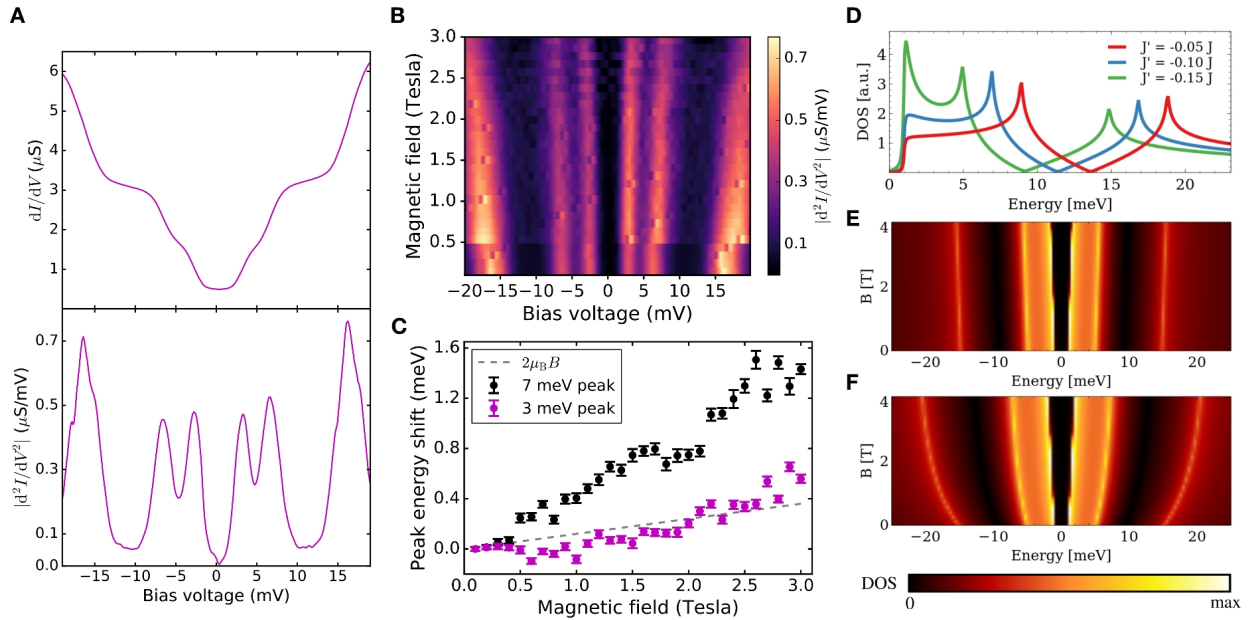

**Fig. 4. (A)** Top panel: Differential conductance versus a DC bias voltage for a bilayer CrI$_3$ barrier device at zero applied magnetic field. The AC excitation was 200 µV and the temperature was 300 mK. Bottom panel: absolute value of d$^2I$/d$V^2$ versus a DC bias voltage, obtained via numerical differentiation of the data in panel (a). According to the theory of inelastic tunneling spectroscopy, the peaks in d$^2I$/d$V^2$ correspond to phonon or magnon excitations of the barrier or electrodes. **(B)** |d$^2I$/d$V^2$| (color scale at right) versus applied magnetic field and DC bias voltage. All three inelastic peaks increase in energy as the applied field is increased. **(C)** Energy of the two lowest energy

inelastic peaks versus applied magnetic field. The zero field energy is subtracted from both peaks for clarity. The dashed grey line shows the Zeeman energy shift of a 2 Bohr magneton magnetic moment (0.12 meV/Tesla), which roughly matches the evolution of the 3 meV peak. **(D)** Calculated magnon density of states (DOS) for $CrI_3$. **(E)** Dispersion of magnons with applied magnetic field at zero temperature. **(F)** Renormalized magnon dispersion with magnetic field at finite temperature.

## Supplementary Materials:

### Materials and Methods:

Our fabrication starts with preparation of Au/Cr wires and bond pads on a 295 nm $SiO_2$/Si wafer. The Au is 33 nm thick and the Cr is 2 nm thick. The wires are prepared by electron beam lithography using bilayer acrylic resist and lift-off processing. The chip with wires is fixed to a ceramic chip carrier using silver paint and wire bonded to the carrier. This assembly is loaded into a glove box for further processing (in which we transfer the heterostructure onto the wires, described below).

The van der Waals assembly of the graphite/$CrI_3$/graphite heterostructure was carried out in an inert argon environment ($H_2O$ and $O_2$ <0.1 ppm). The stack was prepared by sequentially picking up each crystal (top boron nitride, top graphite, $CrI_3$, bottom graphite, and finally bottom boron nitride) from $SiO_2$/Si substrates. We used a stamp-based dry transfer technique similar to those reported elsewhere[35]. For our stamp, we used a poly(bisphenol A carbonate) film stretched over a piece of polydimethylsiloxane.

After the heterostructure was assembled on the stamp, the assembly was pressed onto the Au/Cr wires so that two wires contacted each of the graphite fingers. The heterostructure was released by heating to 170°C. Finally, a glass coverslip was attached to the top of the chip carrier using Crystalbond in order to hermetically seal the device.

Following fabrication, the device was loaded into a helium-3 cryostat. Conductance measurements were carried out using standard low frequency lock-in techniques (the excitation frequency was always <20 Hz). We verified our results using DC measurements where a DC bias was applied to the sample and the resulting DC current was measured with a current preamplifier. For all conductance measurements, unless noted otherwise, the sample temperature was held at either 300 mK or 4.2 K, without strong temperature dependence in this range.

### I. Thermally activated transport in graphite/$CrI_3$/graphite junctions

Figure S1 shows the tunnel conductance for a tetralayer $CrI_3$ barrier as a function of inverse temperature above the magnetic transition, plotted on a semilogarithmic scale. The AC excitation voltage is 5 mV. At higher temperatures the order of magnitude changes linearly with $1/T$, indicating thermally activated transport. The thermal activation gap $\Delta$ is calculated by a fit to:

$$\frac{dI}{dV} = C exp\left(-\frac{\Delta}{2k_B}\frac{1}{T}\right) \qquad (1)$$

where $C$ is a constant and $k_B$ is the Boltzmann constant. Our fit gives a value of approximately 158 meV for this junction, consistent with data from other devices.

### II. Transport through $CrI_3$ barriers in high magnetic fields

In Figure S2A, we show the zero bias conductance of a graphite/bilayer $CrI_3$/graphite junction up to 8 Tesla with an AC excitation of 200 µV. There is a single sharp step in conductance when the external field is increased beyond 0.85 T. At this transition, the junction conductance increases by a factor of about 1.9. There are no steps in the conductance above this transition, consistent with previous magnetization data for bilayer CrI$_3$. We have not observed switching events above 2 T for CrI$_3$ barriers of any thickness. For this device, we detect the onset of Shubnikov-de Haas oscillations above 2 T from the thin graphite contacts. Surprisingly, the zero bias conductance drops to zero as the field is increased. We have further explored this feature as a function of the DC bias and magnetic field, finding that it corresponds to a gap-like feature that opens at high field (Fig. S2B). The origin of this behavior is not presently understood.

### III. Derivation of the magnetoresistance for three and four layer CrI$_3$ barriers

For a trilayer barrier, the total conductance in the aligned configuration is $T_P^3 + T_{AP}^3$, whereas the conductance in the antiparallel configuration is $T_P T_{AP} T_P + T_{AP} T_P T_{AP}$. The ratio of high field to low field conductance is:

$$\frac{G_{\text{HI}}}{G_{\text{LO}}} = \frac{T_P^3 + T_{AP}^3}{T_P^2 T_{AP} + T_{AP}^2 T_P}. \tag{2}$$

Based on a similar calculation, we can calculate the equivalent ratio for a four layer CrI$_3$ barrier:

$$\frac{G_{\text{HI}}}{G_{\text{LO}}} = \frac{T_P^4 + T_{AP}^4}{2 T_P^2 T_{AP}^2}. \tag{3}$$

We use Eq. 2 and 3, together with the formula for bilayer CrI$_3$ given in the main text, to produce the spin filter model fits shown in Fig. 3A of the main text. Note that the above ratios depend only on $T_P/T_{AP}$.

### IV. Junction conductance versus magnetic field for trilayer CrI$_3$ barriers

In Figure S3 we plot the conductance of a graphite/CrI$_3$/graphite junction with a trilayer CrI$_3$ barrier as a function of magnetic field at a temperature of 4.2 K. The AC excitation voltage was 300 µV. The purple and black lines indicate increasing and decreasing direction of the magnetic field sweep, respectively.

### V. Tunneling current and differential conductivity as function of applied bias above 25 meV

The main text reports d$I$/d$V$ and d$^2I$/d$V^2$ data as a function of the applied DC bias $V_{\text{DC}}$. We showed data for a bilayer CrI$_3$ tunnel barrier and for applied bias below 25 meV. We have also measured d$I$/d$V$ versus $V_{\text{DC}}$ for multiple devices and for generator voltages up to 300 meV. This dataset reveals a broader spectrum of inelastic features and the onset of Fowler-Nordheim tunneling around 200 meV.

Figure S4A shows d$I$/d$V$ (top panel) and d$^2I$/d$V^2$ (bottom panel) for a tetralayer CrI$_3$ barrier. The AC excitation was 1 meV and the temperature was 4.2 K. The prominent low bias inelastic features discussed in the main text are visible, along with additional features at higher bias. We have indicated the energies of graphite phonons commonly observed in inelastic tunneling experiments[26,27] as grey dashed lines. Many of the graphite phonon lines are accompanied by apparent features in $|d^2I/dV^2|$, although the 3 meV, 7 meV, and 16 meV peaks are far more prominent. (Note that the 16 meV peak in fact coincides with the $\Gamma$ point ZA phonon of graphite.) There also appears to be a rapid increase in the conductance around 200 meV, which we investigate by pushing to higher DC bias voltages across another CrI$_3$ barrier.

Figure S4B shows the DC current (calculated from the integrated d$I$/d$V$) as a function of the DC voltage for a trilayer CrI$_3$ barrier (black line) at 4.2 K and with an applied magnetic field of 2.5 T. The tunneling current can be estimated using Simmons' model[36]:

$$I(V_{DC}) \propto \int_0^{eV_{DC}} dE\, E\, T(E) + eV_{DC} \int_{eV_{DC}}^{E_F} dE\, T(E) \tag{4}$$

where $T(E)$ is the transmission probability as a function of energy below the Fermi level, $E_F$. For large DC biases, above a few 10s of meV in our parameter range, the second term becomes negligible (and exactly zero for $V_{DC} > E_F/e$). We therefore ignore this term in our fits. We can estimate the transmission probability in the WKB approximation:

$$T(E) = \left| exp\left( -\frac{2\sqrt{2m_*}}{\hbar} \int_0^d dx \sqrt{\Phi - \frac{eV_{DC}x}{d} + E} \right) \right| \tag{5}$$

where $m_*$ is the effective mass of the barrier, $\Phi$ is the barrier height, $\hbar$ is the reduced Planck constant, $e$ is the elementary charge, and $d$ is barrier thickness. The purple line in Fig. S4B is a fit of the measured $I(V)$ curve to Eq. 4 (without the second term) giving $m_* = 1.8$ electron masses and $\Phi = 0.2$ eV, both reasonable values.

As a consistency check, we can also fit the data to the usual Fowler-Nordheim equation (36) strictly valid only for $V_{DC} > \Phi$:

$$\frac{I}{V_{DC}^2} = C exp\left( -\frac{4d\sqrt{2m_*\Phi^3}}{3\hbar e} \frac{1}{V_{DC}} \right) \tag{6}$$

where $C$ is a constant. Figure S4C shows $I/V_{DC}^2$ versus $1/V_{DC}$ on a semilog scale (the x-axis has been restricted to emphasize the high bias behavior). For $V_{DC} > \Phi$ the plot shows the expected linear trend. Linear fits (dashed purple lines) give $4d\sqrt{2m_*\Phi^3}/3\hbar e \approx 1.5$ V, whereas the previously determined effective mass and barrier height gives 1.9 V.

## VI. Derivation of the magnon density of states for monolayer CrI$_3$

In the following we derive an effective Hamiltonian for the spin waves in these system. We start with a Heisenberg Hamiltonian in the honeycomb lattice for the Cr ions:

$$\mathcal{H} = -J \sum_{<ij>} \vec{S_i} \cdot \vec{S_j} - J' \sum_{\ll ij \gg} \vec{S_i} \cdot \vec{S_j} - K \sum_{<ij>} S_i^z S_j^z + B_z \sum_i S_i^z \tag{7}$$

where $J$ is the exchange between nearest neighbors, $J'$ is the exchange between next-nearest neighbors, $K$ is the anisotropic exchange, $<>$ denotes the sum over nearest neighbors, and $\ll \gg$ denotes the sum over next-nearest neighbors. We take $J$, $K > 0$ and $J > K$. This results in a ferromagnetic ground state with easy axis anisotropy along the z axis. Finally, the term *J'* controls the location of the peaks of the magnon spectra.

The magnon spectra of the previous Hamiltonian is obtained by a Holstein-Primakoff transformation, which to second order in bosonic $b_i$ reads:

$$S_i^+ \approx \sqrt{2S}\left(1 - \frac{b_i^\dagger b_i}{4S}\right)b_i; \quad S_i^- \approx \sqrt{2S}b_i^\dagger\left(1 - \frac{b_i^\dagger b_i}{4S}\right); \quad S_i^z = S - b_i^\dagger b_i \tag{8}$$

Substituting the previous formulas in the Heisenberg Hamiltonian we obtain a bosonic Hamiltonian of the form:

$$H = B_z \sum_i b_i^\dagger b_i - \sum_{ij} t_{ij}\, b_i^\dagger b_j + u_{ij} b_i^\dagger b_j b_i^\dagger b_j + v_{ij} b_i^\dagger b_i b_j^\dagger b_j \tag{9}$$

where $t_{ij}$, $u_{ij}$, and $v_{ij}$ are functions of *J*, *J'*, and *K*. The previous Hamiltonian has four field operators so it cannot generically be solved. The Hamiltonian can be made quadratic with the mean field ansatz:

$$b_i^\dagger b_j b_i^\dagger b_j \approx 2\langle b_i^\dagger b_j\rangle b_i^\dagger b_j + \langle b_i^\dagger b_i\rangle b_j^\dagger b_j + \langle b_j^\dagger b_j\rangle b_i^\dagger b_i \tag{10}$$

where we calculate the occupation numbers of the normal modes $b_k$ at low temperature as:

$$\langle b_k^\dagger b_k\rangle = \frac{1}{e^{\beta(\Delta+\rho k^2)} - 1} \approx e^{-\beta(\Delta+\rho k^2)} \tag{11}$$

Approximating $\langle b_i^\dagger b_j\rangle \approx \langle b_i^\dagger b_i\rangle$, we can express the effective magnon hopping in the low temperature limit as:

$$t_{ij}(B_z) = t_{ij}(\infty)\left(1 - C_3 e^{-\beta|B_z|}\right) \tag{12}$$

for the effective mean field spin wave Hamiltonian:

$$H_{MF} = -\sum_{ij} t_{ij}(B_z)\, b_i^\dagger b_j + B_z \sum_i b_i^\dagger b_i \tag{13}$$

This shows that the effect of the off-plane magnetic field is to renormalize the effective magnon hopping, and in particular the effective *g*-factor. The spectrum of the previous Hamiltonian yields the field-dependent magnon density of states. In general, the spectrum shows three peaks: two peaks that can be related to the van Hove singularities of the honeycomb lattice, and a third low energy peak controlled by the next-nearest neighbor exchange. We note that although the next-nearest neighbor exchange is antiferromagnetic, the combination of nearest neighbor exchange and anisotropic exchange will impose a ferromagnetic ground state for small values of the next-nearest neighbor exchange. The exchange constants of the parent Heisenberg Hamiltonian were chosen so that the spin wave peaks at located at the energies found in the experiment.

**VII. Magnetic field dependence of the inelastic tunneling spectra in additional devices**

We have also found variations among devices in the magnetic field dependence: for one tetralayer $CrI_3$ barrier we observed energy shifts close to $2\mu_B B$, shown in Fig. S5.

## VIII. Density functional theory methods

We performed density functional theory calculations with Quantum Espresso[37] using PAW pseudopotentials and GGA exchange correlation functional. We used a geometry consisting of three layers of $CrI_3$ and three layers of graphite, consisting of a 3 by 3 unit cell of graphite with periodic boundary conditions in the $z$ direction and ABC stacking. Structural relaxations were performed with spin polarization. Band structure calculations show that the majority *e.g.* $CrI_3$ bands are located close to the charge neutrality point of graphite, and we expect a very small charge transfer from graphite to $CrI_3$. Minority spin bands reside in the gap of $CrI_3$, and are located 1 eV away from minority $t_{2g}$ bands of $CrI_3$. We also computed heterostructures made of monolayer graphene on top of monolayer $CrI_3$ and monolayer graphene on top of bilayer $CrI_3$. In all instances the charge neutrality point is located in the majority *e.g.* $CrI_3$ bands.

## IX. Temperature dependence of the inelastic tunneling features

We measured the differential conductance as a function of both temperature and applied bias for a tetralayer $CrI_3$ barrier junction. This was done by heating the sample to about 65 K, and letting it cool slowly while sweeping the DC offset voltage back and forth. The temperature decrease during the voltage sweep was always less than 2 K. We applied a magnetic field of 1.8 T during the measurements to keep the sample in the fully aligned magnetic configuration. This makes the data easier to interpret and collect since the conductance has a strong temperature dependence in the antiferromagnetic state. We then took the numerical derivative to obtain $|d^2I/d^2V|$ as a function of temperature and applied bias, plotted in Fig. S6B. The dashed white line indicates the magnetic transition at 51 K, as determined from the temperature dependence of the resistance in the antiferromagnetic state (Fig. S6A).

The sharp inelastic features at low temperature quickly evolve into a single broad feature, which then decreases in intensity, disappearing completely just below the magnetic transition. We can quantify this trend by plotting the total IETS spectral weight (implemented by summing the IETS spectra over the range -20 meV to 20 meV). This is plotted in Fig. S6C, where we see that the IETS signal begins to onset just below 50 K. We do not expect a complete loss of spectral weight at this temperature from thermal broadening alone[38], suggesting some relation between the onset of magnetism and the inelastic tunneling features.

## X. Bias dependence of the magnetoresistance ratio

In the main text we have focused on the magnetoresistance ratio at zero bias, but it also shows an interesting bias dependence. In Figure S7A we plot, for a tetralayer $CrI_3$ barrier, the ratio of the differential conductance taken with an external field of 2.5 T (in the fully aligned state) and -0.2 T (in the antiferromagnetic state). The AC excitation was 500 µV RMS, and the temperature was 4.2 K. We see that, although the differential magnetoresistance ratio remains high for all bias values, it has a peak around zero bias. This has been observed in other magnetic tunnel junctions and could be a product of spin flip magnon scattering. For Figure S7A the AC excitation frequency was reduced until the in-phase component of the tunneling current saturated, giving the correct quasi-DC value of the magnetoresistance ratio.

Figure S7B shows an analogous measurement for a $CrI_3$ bilayer tunnel barrier. Here, the temperature was 300 mK and the AC excitation was 200 µV. The data for the ferromagnetic configuration was taken with an applied magnetic field of 0.6 T and the data for the antiferromagnetic configuration was taken with 0.1 T applied magnetic field. This data also shows a sharp decrease in differential magnetoresistance ratio around zero bias, but in this case the differential magnetoresistance ratio drops as low as 10% to 20%. This indicates that electrons tunneling through bilayer $CrI_3$ with energy above a few meV are likely not strongly spin polarized.

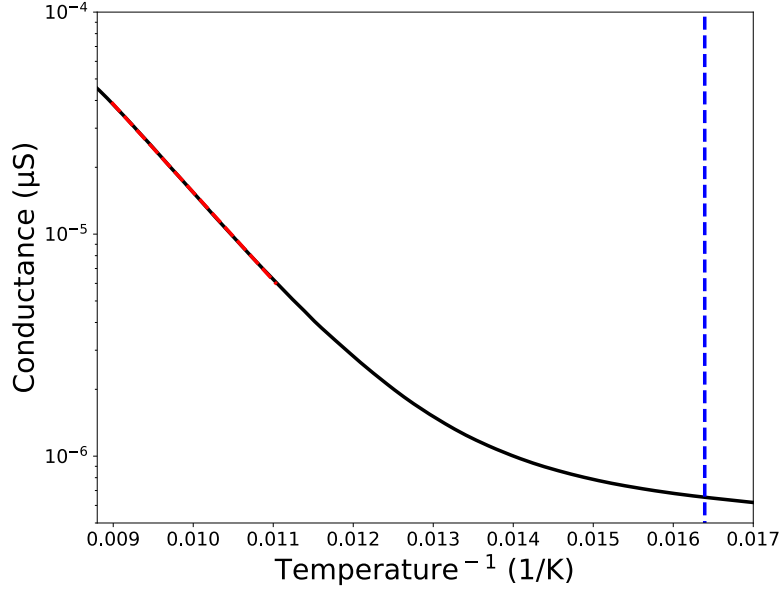

**Fig. S1.** Arrhenius plot of zero bias junction conductance vs. inverse temperature for a graphite/tetralayer CrI$_3$/graphite junction cooled without an applied magnetic field. The red line indicates the linear fit used to extract the thermal activation gap above the metamagnetic transition. The bulk Curie temperature (61 K) is shown with the blue dashed line.

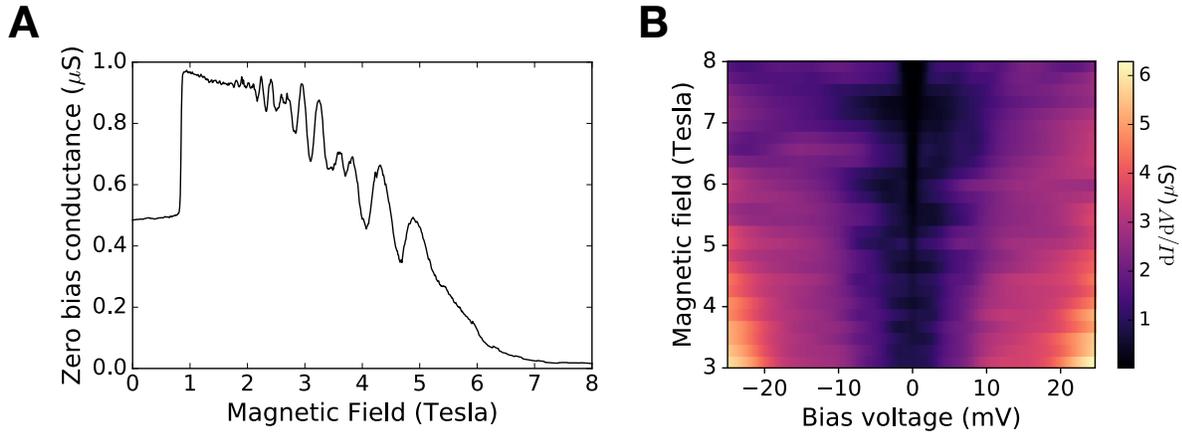

**Fig. S2. (A)** Conductance through a bilayer CrI$_3$ tunnel barrier as a function of an out-of-plane applied magnetic field with 200 μV AC excitation. The external field is swept up from zero. Beyond 2 Tesla, we observe Shubnikov-de Haas oscillations in conductance due to the graphite contacts in the device. **(B)** Differential conductance as a function of DC bias and applied magnetic field. In addition to the linearly dispersing inelastic features, Shubnikov-de Haas and a pronounced gap opening at high field are visible.

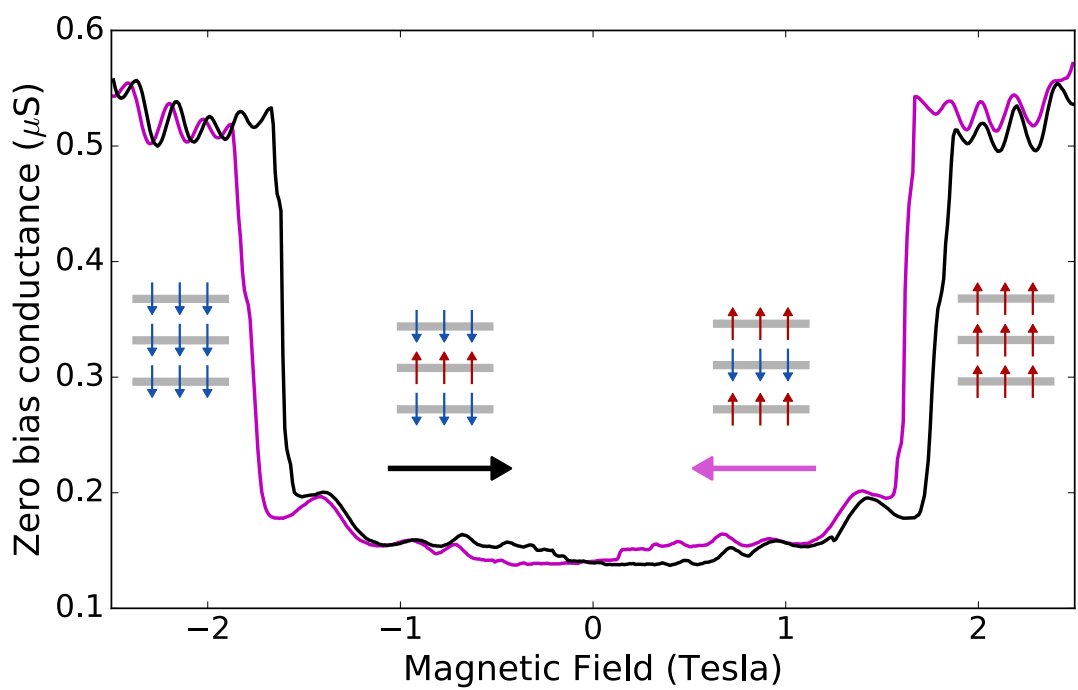

**Fig S3.** Magnetic field dependence of the junction conductance for a graphite/CrI$_3$/graphite junction with a trilayer CrI$_3$ barrier, at a temperature of 4.2 K. The black and purple lines represent up and down sweeps of the magnetization, respectively. The AC excitation voltage was 300 μV.

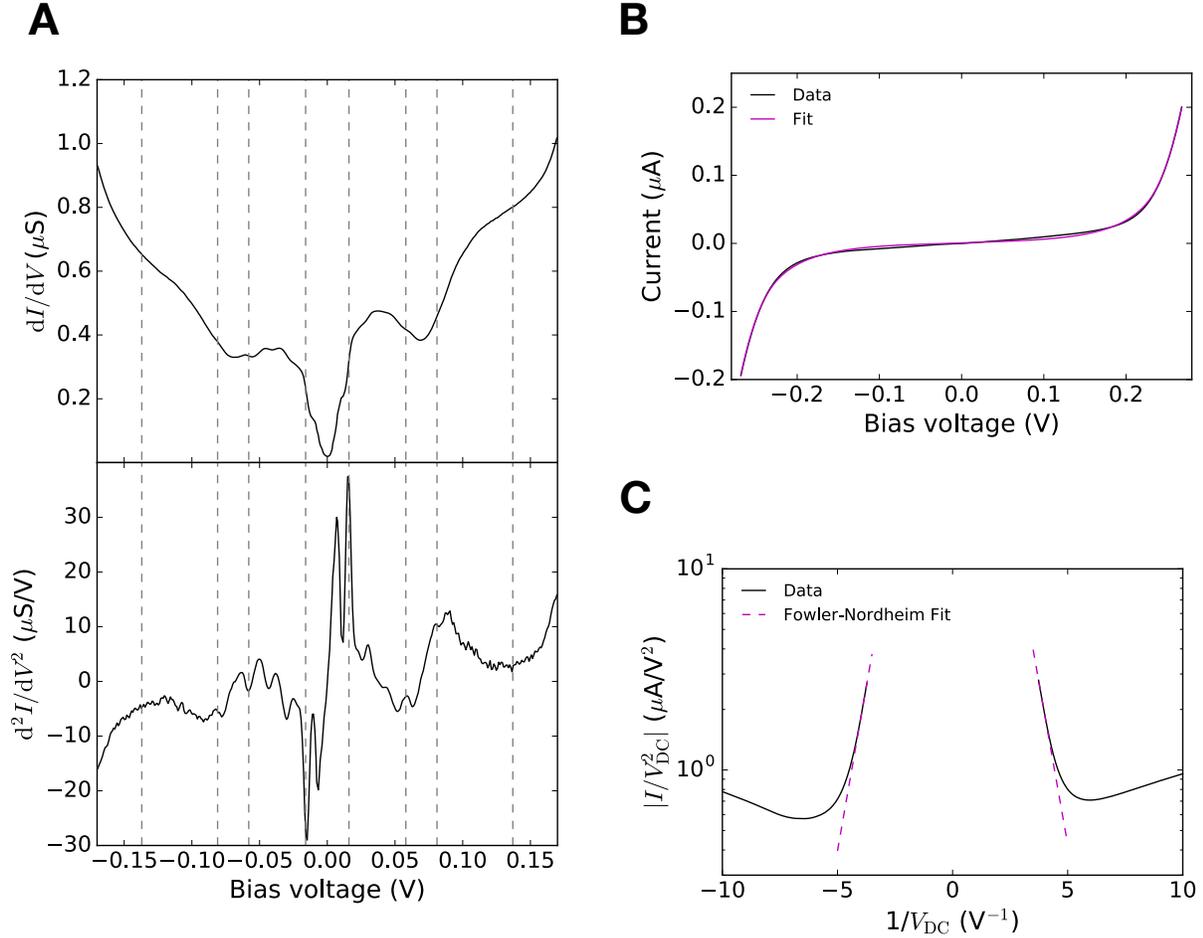

**Fig. S4.** **(A)** Top panel: Differential conductance through a tetralayer CrI$_3$ tunnel barrier as a function of DC bias. The temperature was 4.2 K and the AC excitation was 1 meV. Bottom panel: $|d^2I/dV^2|$ as a function of DC bias calculated as the numerical derivative of the data in top panel. The grey lines indicate the energies of graphite phonons that have been observed in previous inelastic tunneling experiments. **(B)** Current versus DC bias (black line) for a trilayer CrI$_3$ tunnel barrier at 4.2 K and 2.5 T magnetic field. The purple line is a fit to Eq. 3. **(C)** $I/V_{DC}^2$ versus $1/V_{DC}$ on a semilog scale (black line). The data is the same as panel (B), but over a restricted bias range. The dashed purple lines are fits of the high bias data to the Fowler-Nordheim formula (Eq. 5).

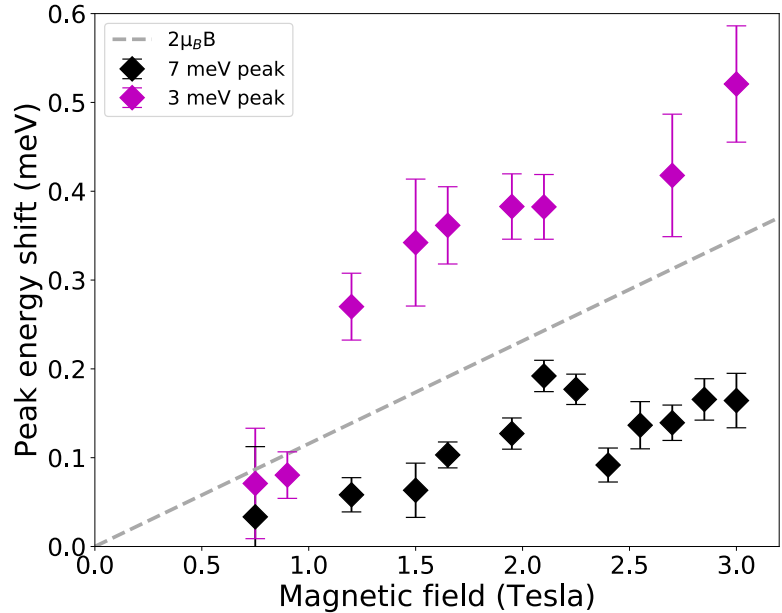

**Fig. S5.** Shift in peak energy with applied magnetic field for the innermost two peaks of the IETS for a tetralayer $CrI_3$ junction. The zero field peak values were subtracted for clarity. The grey dashed line represents the Zeeman energy shift for a magnetic moment of 2 Bohr magnetons (roughly 0.12 meV/Tesla).

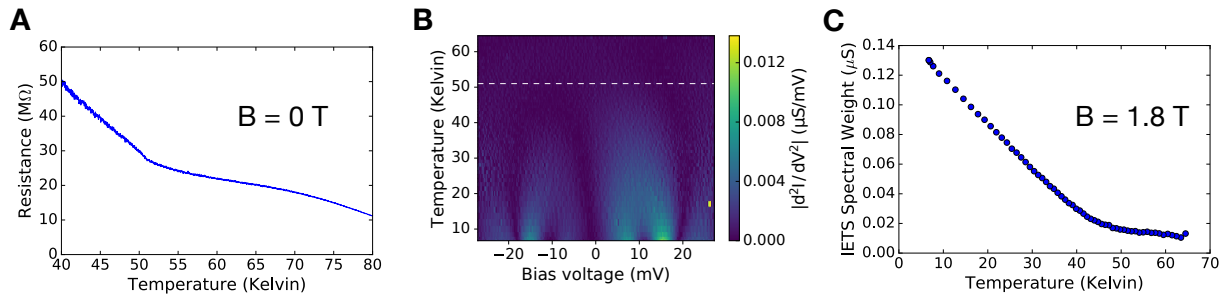

**Fig S6. (A)** Temperature dependence of the resistance for a tetralayer $CrI_3$ device, taken without applied magnetic field. The onset of the magnetoresistance effect (appearing as a kink) gives the Neel temperature, 51 K, of this sample. **(B)** $|d^2I/dV^2|$ for the same device, plotted as a function of the DC bias voltage and temperature. The dashed white line is at 51 K. The applied magnetic field for this data was 1.8 T, to avoid artifacts from the large resistance increase at zero magnetic field. **(C)** IETS data integrated from -20 meV to 20 meV, showing the onset of inelastic tunneling below 50 K. The error bars are significantly smaller than the marker size.

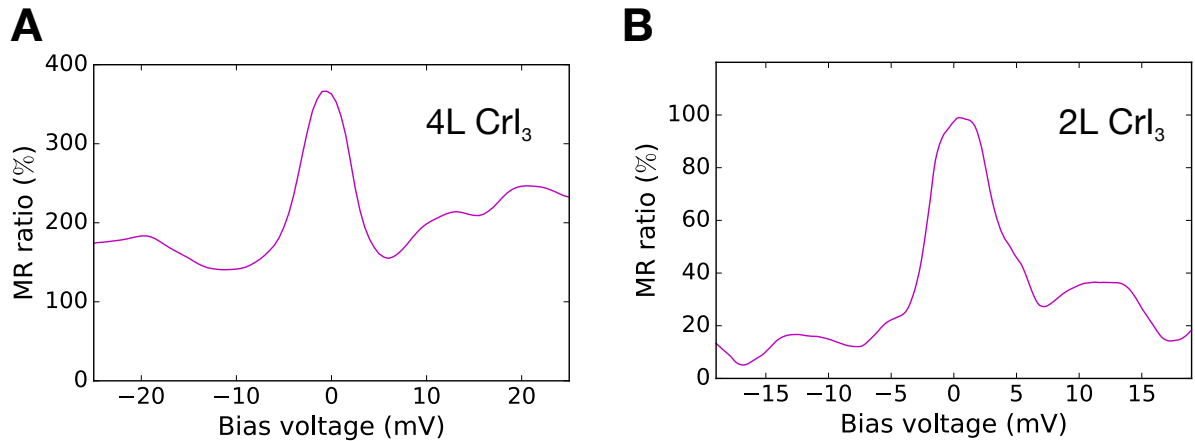

**Fig. S7. (A)** Ratio of differential conductance in the ferromagnetic configuration (with an applied field of 2.5 T) to differential conductance in the antiferromagnetic configuration (with an applied field of -0.2 T) as a function of a DC bias voltage. This data was taken on a tetralayer $CrI_3$ barrier junction. The temperature was 4.2 K, and the AC excitation was 500 µV RMS. **(B)** Ratio of differential conductance in the ferromagnetic configuration (with an applied field of 0.6 T) to differential conductance in the antiferromagnetic configuration (with an applied field of 0.1 T) as a function of a DC bias voltage. This data was taken on a bilayer $CrI_3$ barrier junction. The temperature was 300 mK, and the AC excitation was 200 µV RMS.